\documentclass[12pt]{article}
\begin{document}
\begin{center}
{\large\bf Inhomogeneous Cosmology, Inflation and Late-Time
Accelerating Universe} \vskip 0.3 true in {\large J. W. Moffat}
\vskip 0.3 true in {\it The Perimeter Institute for Theoretical
Physics, Waterloo, Ontario, N2L 2Y5, Canada} \vskip 0.3 true in
and \vskip 0.3 true in {\it Department of Physics, University of
Waterloo, Waterloo, Ontario N2L 3G1, Canada}
\end{center}
\begin{abstract}%
An exact inhomogeneous solution of Einstein's field equations is
shown to be able to inflate in a non-uniform way in the early
universe and explain anomalies in the WMAP power spectrum data. It
is also possible for the model to explain the accelerated
expansion of the universe by late-time inhomogeneous structure.
\end{abstract}
\vskip 0.2 true in e-mail: john.moffat@utoronto.ca


\section{Introduction}

In the following, we shall investigate an exact inhomogeneous
cosmological solution of Einstein's field equations obtained by
Szekeres~\cite{Szekeres}, for an irrotational dust dominated
universe and subsequently generalized by Szafron~\cite{Szafron} and
Szafron and Wainright~\cite{Wainright,Krasinski} to the case when
the pressure $p$ is non--zero. The cosmological model contains the
Friedmann-Lema\^{i}tre-Robertson-Walker (FLRW) cosmology and the
spherically symmetric Lema\^{i}tre-Tolman-Bondi (LTB)
model~\cite{Lemaitre,Tolman,Bondi} as special solutions. The Szafron
model describes a more general inhomogeneous spacetime with no
prescribed initial symmetry~\cite{Bonnor}.

We will show that the Szafron inhomogeneous cosmology in the early
universe near the Planck time can lead to an inhomogeneous
inflationary period, which can predict that the primordial power
spectrum is not uniform across the sky. Recent investigations have
revealed that there appears to be a lack of power in the CMB power
spectrum above $\sim 60^0$ and anisotropy in hot and cold spots on
the sky~\cite{Starkman,Eriksen,Tegmark,Starkman2,Land}. Moreover,
there has been the claim that there is a peculiar alignment
between the quadrupole and octopole moments called the ``axis of
evil''~\cite{Land}. These anomalies do not agree with the standard
homogeneous and isotropic inflationary
models~\cite{Guth,Linde,Albrecht,Lyth}. The anomalies in the WMAP
data could be explained by the inhomogeneous solution presented in
the following. Several possible explanations for the anomalies in
the WMAP data have been proposed including an exact planar
solution of Einstein's field equations~\cite{Buniy}, the violation
of Lorentz symmetry and rotational
invariance~\cite{Wise,Kostelecky,Moffat,Moffat2} and a cutoff of
inflation that leads to a loss of primordial power
spectrum~\cite{Contaldi}.

The problem of explaining the acceleration of the universe as
determined by supernovae data and the cosmic microwave background
(CMB) data is one of the most significant outstanding problems in
modern physics and
cosmology~\cite{Perlmutter,Riess,Hinshaw,Spergel}. The standard
explanation is either based on postulating a cosmological constant
$\Lambda$ or assuming that some form of uniform dark energy with
negative pressure exists in the universe~\cite{Peebles}. The
explanation as to why the cosmological constant is zero or very
small has led to a crisis in physics and cosmology.

The LTB model has been used to explain the non-Gaussian behavior
observed in the WMAP data~\cite{Hinshaw,Spergel} and give a
possible explanation for the late-time acceleration of the
universe~\cite{Moffat3,Moffat4,Moffat5,Celerier}. A criticism of
the LTB model is that it assumes a spherically symmetric universe
with one spatial degree of inhomogeneity, requiring a center of
the universe and that observers be located not too far from the
center to avoid undetected large anisotropy. With this restriction
the LTB solution has provided a toy-model description of the
late-time, non-linear regime with voids and collapsing matter.

\section{Inhomogeneous Cosmological Solution}

The metric takes the form\footnote{We use units with the speed of
light c=1.}
\begin{equation}
\label{metric}
ds^2=dt^2-R^2(\mathbf{x},t)(dx^2+dy^2)-S^2(\mathbf{x},t)dz^2,
\end{equation}
where $R(\mathbf{x},t)$ and $S(\mathbf{x},t)$ are to be determined
from Einstein's field equations:
\begin{equation}
G_{\mu\nu}\equiv R_{\mu\nu}-\frac{1}{2}g_{\mu\nu}R=8\pi
GT_{\mu\nu}+\Lambda g_{\mu\nu},
\end{equation}
where $T_{\mu\nu}$ denotes the perfect fluid energy-momentum
tensor
\begin{equation}
T_{\mu\nu}=(\rho+p)u_\mu u_\nu-pg_{\mu\nu},
\end{equation}
and $\Lambda$ is the cosmological constant. Moreover, $\rho$
denotes the energy-density of matter, $p$ the pressure and $u_\mu$
the velocity field of the fluid which is normalized to $u^\mu
u_\mu=1$. The coordinates are assumed to be comoving so that
$u^\mu=\delta^\mu_0$ and ${\dot u}^\mu=0$ where ${\dot
u^\mu}=du^\mu/dt$. Let us introduce the notation
\begin{equation}
R(\mathbf{x},t)=\exp(\beta(\mathbf{x},t)),
\end{equation}
and
\begin{equation}
S(\mathbf{x},t)=\exp(\alpha(\mathbf{x},t)).
\end{equation}

Szafron~\cite{Szafron,Wainright,Krasinski} solved the Einstein
equations with $p\not= 0$ and $\Lambda=0$, generalizing the dust
solution of Szekeres~\cite{Szekeres}. There are two classes of
solution $\beta'\not=0$ and $\beta'=0$ where
$\beta'=\partial\beta/\partial z$. The more general solution
$\beta'\not= 0$ is given for $\Lambda=0$ by
\begin{equation}
\label{Req}
R(\mathbf{x},t)\equiv\exp(\beta(\mathbf{x},t))=a(z,t)\exp(\nu(\mathbf{x})),
\end{equation}
\begin{equation}
\label{Seq}
 S(\mathbf{x},t)\equiv\exp(\alpha(\mathbf{x},t))
=h(z)\exp(-\nu(\mathbf{x}))
\partial\exp(\beta(\mathbf{x},t)/\partial z,
\end{equation}
where
\begin{equation}
\label{nueq}
\exp(-\nu(\mathbf{x}))=A(z)(x^2+y^2)+2B_1(z)x+2B_2(z)y+C(z).
\end{equation}
The $A(z), B_1(z), B_2(z), C(z)$ and $h(z)$ are arbitrary functions
of $z$. Moreover, $a(z,t)$ satisfies the equation
\begin{equation}
\label{Friedmannz} 2\frac{{\ddot a(z,t)}}{a(z,t)}+\frac{{\dot
a^2(z,t)}}{a^2(z,t)}+\frac{k(z)}{a^2(z,t)} =-8\pi Gp(t),
\end{equation}
which has the same form as one of the Friedmann equations in FLRW
cosmology, except that $a=a(z,t)$ and $k=k(z)$.  The function $k(z)$
is determined by
\begin{equation}
k(z)=4\bigg(A(z)C(z)-B^2_1(z)-B^2_2(z)-\frac{1}{4}\frac{1}{h^2(z)}\biggr).
\end{equation}
The case $\beta'\rightarrow 0$ is singular. Eq.(\ref{Friedmannz})
can be formally integrated once $p(t)$ is specified. The density
equation is given by
\begin{equation}
\label{denseq}
\ddot{\alpha}(\mathbf{x},t)+2\ddot{\beta}(\mathbf{x},t)+\dot{\alpha}^2(\mathbf{x},t)
+2\dot{\beta}^2(\mathbf{x},t)=-4\pi G(\rho(\mathbf{x},t)+3p(t)).
\end{equation}

An algorithm for generating an exact solution is to specify
explicitly $p=p(t)$ and solve (\ref{Friedmannz}) for $a(z,t)$. The
metric is now obtained by solving for $R(\mathbf{x},t)$ and
$S(\mathbf{x},t)$ from Eqs.(\ref{Req}), (\ref{Seq}) and
(\ref{nueq}). The equation for the density $\rho(\mathbf{x},t)$ is
given by (\ref{denseq}).

We shall consider, in the following, the simpler Szafron solution
with $\beta'=0$ given by
\begin{equation}
\label{Requation}
R(\mathbf{x},t)\equiv\exp(\beta(\mathbf{x},t))=\frac{a(t)}{1+\frac{1}{4}k(x^2+y^2)},
\end{equation}
\begin{equation}
\label{Sequation}
S(\mathbf{x},t)\equiv\exp(\alpha(\mathbf{x},t))=\lambda(z,t)+a(t)\Sigma(\mathbf{x}),
\end{equation}
\begin{equation}
\Sigma(\mathbf{x})=\frac{\frac{1}{2}U(z)(x^2+y^2)+V_1(z)x+V_2(z)y+2W(z)}{1+\frac{1}{4}k(x^2+y^2)}.
\end{equation}
Now $k$ is a constant, $U(z), V_1(z), V_2(z)$ and $W(z)$ are
arbitrary functions of $z$ and $a(t)$ is determined by the Friedmann
equation
\begin{equation}
\label{pressureeq} 2\frac{{\ddot a(t)}}{a(t)}+\frac{{\dot
a}^2(t)}{a^2(t)}+\frac{k}{a^2(t)}=-8\pi Gp(t).
\end{equation}
We can choose without loss of generality $W(z)=0$ and
$\lambda(z,t)$ is determined by
\begin{equation}
\label{lambdaeq} {\ddot\lambda(z,t)}a(t)+{\dot\lambda(z,t)}{\dot
a(t)}+\lambda(z,t){\ddot a(t)}=-8\pi G\lambda(z,t)a(t)p(t)+U(z).
\end{equation}
The matter density equation is given by
\begin{equation}
\label{mattereq}
-\frac{2}{3}\biggl[{\ddot\lambda(z,t)}-\lambda(z,t)\frac{\ddot
a(t)}{a(t)}\biggr]\exp(-\alpha(\mathbf{x}))+H^2(t)+\frac{k}{a^2(t)}
=\biggl(\frac{8\pi G}{3}\biggr)\rho(\mathbf{x},t),
\end{equation}
where $H(t)={\dot a}(t)/a(t)$. Eqs.(\ref{pressureeq}) and
(\ref{lambdaeq}) can be solved once the pressure $p=p(t)$ is
specified, while (\ref{mattereq}) determines the density
$\rho(\mathbf{x},t)$. The FLRW spacetime is obtained when
$\lambda(z,t)=U(z)=0$. When $U(z)=0$ and $V_1(z)=V_2(z)=0$, the
model possesses a 3-dimensional symmetry group acting on
2-dimensional orbits. The symmetry is spherical, plane or
hyperbolic when $k >0, k=0$ or $k < 0$, respectively.

When $k=0$ the line element becomes
\begin{equation}
ds^2=dt^2-a^2(t)(dx^2+dy^2)-S^2(\mathbf{x},t)dz^2.
\end{equation}
A cosmological solution is obtained for $a(t)$ and
$S(\mathbf{x},t)$ when $p(t)$ and $\rho(\mathbf{x},t)$ are
specified.

\section{Coarse-Grained Spatial Averaging of Inhomogeneous Model}

For our exact inhomogeneous cosmological model, we are required to
carry out a volume averaging of physical quantities. We define for
a scalar quantity $\Psi$ a coarse-grained spatial
smoothing~\cite{Ellis,Buchert,Kolb}:
\begin{equation}
\langle\Psi(\mathbf{x},t)\rangle_D=\frac{1}{{\cal V_D}}\int_D
d^3x\sqrt{\gamma}\Psi(\mathbf{x},t),
\end{equation}
where
\begin{equation}
{\cal V}_D=\int_Dd^3x\sqrt{\gamma}
\end{equation}
is the volume of the simply-connected domain, $D$, in a
hypersurface. We can define effective scale-factors for our
spatially averaged cosmological model:
\begin{equation}
\label{smoothR} R_D(t)=\langle
R(\mathbf{x},t)\rangle_D=\biggl(\frac{{\cal V}(t)_{RD}}{{\cal
V}_{iD}}\biggr)^{1/3},
\end{equation}
\begin{equation}
S_D(t)=\langle S(\mathbf{x},t)\rangle_D=\biggl(\frac{{\cal
V}(t)_{SD}}{{\cal V}_{iD}}\biggr)^{1/3},
\end{equation}
where ${\cal V}_{iD}$ is the initial spatial volume.

We define the spatially averaged Hubble parameters
\begin{equation}
H_{RD}(t)=\frac{\dot {R}_D(t)}{R_D(t)},\quad H_{SD}(t)=\frac{\dot
{S}_D(t)}{S_D(t)}
\end{equation}
and the effective Hubble expansion parameter
\begin{equation}
H_{\rm eff}(t)=\frac{1}{3}(2H_{RD}(t)+H_{SD}(t)).
\end{equation}

Consider a congruence of curves with a time-like unit vector $V_\mu$
with $g^{\mu\nu}V_\mu V_\nu=1$ and $dV^\mu/ds=\nabla_\nu V^\mu
V^\nu$ is the acceleration of the flow lines. Here, $\nabla_\mu$ and
$s$ denote the covariant derivative with respect to $g_{\mu\nu}$ and
the proper time, respectively. The metric tensor ${h^\mu}_\nu$ is
given by
\begin{equation}
{h^\mu}_\nu={\delta^\mu}_\nu+V^\mu V_\nu
\end{equation}
and describes the metric that projects a vector into its
components in the subspace of the vector tangent space that is
orthogonal to $V$.

We define the vorticity tensor
\begin{equation}
\omega_{\mu\nu}=\nabla_{[\nu} V_{\mu]},
\end{equation}
and the shear tensor
\begin{equation}
\sigma_{\mu\nu}=\theta_{\mu\nu}-\frac{1}{3}\theta h_{\mu\nu},
\end{equation}
where
\begin{equation}
\theta_{\mu\nu}=\nabla_{(\nu}V_{\mu)}.
\end{equation}
The covariant derivative of $V$ can be expressed as
\begin{equation}
\nabla_\nu
V_\mu=\omega_{\mu\nu}+\sigma_{\mu\nu}+\frac{1}{3}h_{\mu\nu}\theta
-\frac{dV_\mu}{ds}V_\nu.
\end{equation}
The volume expansion is given by
\begin{equation}
\label{volumeexpand} \theta\equiv
h^{\mu\nu}\theta_{\mu\nu}=\nabla_\mu V^\mu.
\end{equation}

The volume averaging of the scalar $\Psi$ does not commute with
its time evolution~\cite{Ellis,Buchert,Kolb}:
\begin{equation}
\langle\dot{\Psi}(\mathbf{x},t)\rangle_D-\partial_t\langle\Psi(\mathbf{x},t)\rangle_D
=\langle\Psi(\mathbf{x},t)\rangle_D\langle\theta(\mathbf{x},t)\rangle_D
-\langle\Psi(\mathbf{x},t)\theta(\mathbf{x},t)\rangle_D.
\end{equation}
For inhomogeneous cosmology, the smoothing due to averaging of the
Einstein field equations does not commute with the time evolution
of the non-linear field equations. This leads to extra
contributions in the effective, averaged Einstein field equations,
which do not satisfy the usual energy conditions even though they
are satisfied by the original energy-momentum tensor. It is the
lack of commutativity of the time evolution of the expansion of
the universe in a local patch inside our Hubble horizon, that
circumvents the no-go theorem based on the local Raychaudhuri
equation~\cite{Raychaudhuri,Hawking}, namely, that the expansion
of the universe cannot accelerate when the weak and strong energy
conditions: $\rho > 0$ and $\rho+3p > 0$ are satisfied.

\section{Inhomogeneous Inflationary Model and Microwave Background}

Let us now consider the very early universe and for simplicity
$k=0$. We choose
\begin{equation}
p=-Z={\rm const.},
\end{equation}
where $Z > 0$. Then, a de Sitter solution of Eq.(\ref{pressureeq})
is given by
\begin{equation}
a(t)=\exp(H_at),\quad H_a=\sqrt{\frac{8\pi GZ}{3}}.
\end{equation}
We have from Eqs.(\ref{lambdaeq}) and (\ref{mattereq}):
\begin{equation}
{\ddot\lambda}(z,t)+{\dot\lambda}(z,t)H_a+\lambda(z,t)H_a^2=8\pi
GZ\lambda(z,t)+\exp(-H_at)U(z),
\end{equation}
and
\begin{equation}
\lambda(z,t)H_a^2-{\ddot\lambda}(z,t)+\frac{3}{2}H_a^2\exp(\alpha(\mathbf{x}))
=4\pi G\rho(\mathbf{x},t)\exp(\alpha(\mathbf{x})).
\end{equation}
We now obtain from (\ref{Requation}) and (\ref{Sequation}) the
metric
\begin{equation}
\label{inhomometric}
ds^2=dt^2-\exp(2H_at)(dx^2+dy^2)-[\lambda(z,t)+\exp(H_at)\Sigma(\mathbf{x})]^2dz^2.
\end{equation}

We see that as $t\rightarrow\infty$ the universe inflates in the
$x-y$ plane but could for $\Sigma\sim 0$ be non-inflating along the
$z$ direction. For $\lambda=0$ and $\Sigma=1$, we regain the
inflating de Sitter spacetime metric
\begin{equation}
ds^2=dt^2-\exp(2H_at)(dx^2+dy^2+dz^2).
\end{equation}

When inflation ends, the metric in the early radiation dominated
universe should approach a homogeneous and isotropic FLRW metric as
$t$ increases with $R(\mathbf{x},t)\sim a(t)$ and the equation of
state $p(t)=\rho(t)/3$. After decoupling the FLRW spacetime is the
matter dominated solution with zero pressure, $p(t)=0$.

Let us investigate how the inhomogeneous solution can give rise to a
non-uniform power spectrum. The power spectrum for homogeneous and
isotropic inflation is given by
\begin{equation}
\langle\delta(\mathbf{k}),\delta(\mathbf{q})\rangle
=P(k)\delta^3(\mathbf{k}-\mathbf{q}),
\end{equation}
where $\delta(\mathbf{k})$ denotes the primordial density contrast
and the translational invariance of the inflationary epoch leads
to the modes with different wave numbers being uncoupled. Let us
assume that there exists a vector in the anisotropic direction of
unit vector $\mathbf{n}$. We assume parity symmetry
$\mathbf{k}\rightarrow -\mathbf{k}$ and denote by $\tilde{P}(k)$
the power spectrum caused by the anisotropy of primordial
spacetime. We have~\cite{Wise}:
\begin{equation}
\tilde{P}(\mathbf{k})=P(k)\biggl(1+f(k)(\hat{\mathbf{k}}\cdot\mathbf{n})^2\biggr),
\end{equation}
where $k=\vert\mathbf{k}\vert$, $\hat{\mathbf{k}}$ is the unit
vector along the direction of $\mathbf{k}$ and we have kept only the
lowest power of $\hat{\mathbf{k}}\cdot\mathbf{n}$. We require that
the power spectrum be scale invariant, $P(k)\sim 1/k^3$, with $f(k)$
being independent of $k$, so that $f(k)\sim {\cal A}$ where ${\cal
A}$ is a constant. We now have
\begin{equation}
\tilde{P}(\mathbf{k})=P(k)\biggl(1+{\cal
A}(\hat{\mathbf{k}}\cdot\mathbf{n})^2\biggr).
\end{equation}

The inhomogeneous spacetime gives rise to correlations between
multipole moments that normally are zero for homogeneous and
isotropic inflation. Deviations from homogeneous and isotropic
inflation can be parameterized by
\begin{equation}
\epsilon_I=\biggl(\frac{a(t)-R_D(t)}{K(t)}\biggr),
\end{equation}
where $a(t)\sim \exp(H_at)$, $R_D(t)$ is the smoothed-out scale
factor (\ref{smoothR}) and $K(t)$ is the averaged cosmic scale:
\begin{equation}
K(t)=\frac{1}{3}(2R_D(t)+a(t)).
\end{equation}

\section{Acceleration of the Late-Time Matter Dominated Universe}

Let us now consider the matter dominated solution with zero
pressure, $p=0$. We have
\begin{equation}
\label{matterdom} 2\frac{{\ddot a(t)}}{a(t)}+\frac{{\dot
a}^2(t)}{a^2(t)}+\frac{k}{a^2(t)}=0.
\end{equation}
This equation has the standard Friedmann solution for $k=0$:
\begin{equation}
a(t)=\biggl(\frac{t}{t_0}\biggr)^{2/3}.
\end{equation}
The matter equation is given by (\ref{mattereq}) where
\begin{equation}
\exp(-\alpha(\mathbf{x}))=
\biggl[\lambda(z,t)+\biggl(\frac{t}{t_0}\biggr)^{2/3}\Sigma(\mathbf{x})\biggr]^{-1}.
\end{equation}

Let us define
\begin{equation}
\Omega_m(\mathbf{x},t)=\frac{8\pi G\rho(\mathbf{x},t)}{3H^2(t)},
\end{equation}
and for a spatially flat universe with $k=0$:
\begin{equation}
\Omega_X(\mathbf{x},t)=\frac{2\biggl[{\ddot\lambda(z,t)}-\lambda(z,t)\frac{\ddot
a(t)}{a(t)}\biggr]\exp(-\alpha(\mathbf{x}))}{3H^2(t)}.
\end{equation}
We have
\begin{equation}
\label{densityfrac} \Omega_m(\mathbf{x},t)+\Omega_X(\mathbf{x},t)=1.
\end{equation}

We define spatially averaged $\Omega_m$ and $\Omega_X$:
\begin{equation}
\Omega_{mD}(t)=\langle\Omega_m(\mathbf{x},t)\rangle_D=\frac{1}{{\cal
V_D}}\int_D d^3x\sqrt{\gamma}\Omega_m(\mathbf{x},t),
\end{equation}
\begin{equation}
\Omega_{XD}(t)
=\langle\Omega_X(\mathbf{x},t)\rangle_D=\frac{1}{{\cal V_D}}\int_D
d^3x\sqrt{\gamma}\Omega_X(\mathbf{x},t),
\end{equation}
It follows from (\ref{densityfrac}) that
\begin{equation}
\Omega_{mD}(t)+\Omega_{XD}(t)=1.
\end{equation}
We obtain when $\lambda=0$, the standard FLRW density relation for
an Einstein-de Sitter universe:
\begin{equation}
\Omega_m(t)=1,
\end{equation}
where
\begin{equation}
\Omega_m(t)=\frac{8\pi G}{3}\rho(t).
\end{equation}

By choosing the parameterization:
\begin{equation}
\Omega^0_{mD}=0.28,\quad \Omega^0_{XD}=0.72,
\end{equation}
where $\Omega^0_{mD}$ and $\Omega^0_{XD}$ denote the present
matter and inhomogeneity densities, we can fit the WMAP
data~\cite{Spergel} and the supernovae
data~\cite{Perlmutter,Riess}.

\section{Conclusions}

We have derived a primordial inhomogeneous inflationary model from
an exact inhomogeneous solution of Einstein's field equations.
Because the model has a preferred direction of anisotropic
inflation, it is possible to explain detected anomalies in the
WMAP power spectrum such as the apparent alignment of the CMB
multipoles on very large scales and the loss of power for
$\theta\geq 60^0$. The observed CMB temperature anisotropies can
give a window on the primordial inflationary era. We have given
generic predictions expected from the existence of a preferred
inhomogeneous direction during inflation. The inhomogeneous
inflation should reveal itself in a scale-invariant manner, and
lead to predictions for the primordial fluctuation correlations
determined by a single amplitude ${\cal A}$ and a unit vector
$\mathbf{n}$ signifying a preferred anisotropic direction on the
sky.

A possible explanation of the accelerated expansion of the
universe at late times and the WMAP data is attributed to the
observed late-time non-linear, inhomogeneous galaxy and void
structure. Our exact inhomogeneous solution avoids the
anti-Copernican assumption of a center to the universe, and it
also avoids the postulates of an ill-understood cosmological
constant and ``dark energy''.

\vskip 0.2 true in {\bf Acknowledgments} \vskip 0.2 true in This
work was supported by the Natural Sciences and Engineering Research
Council of Canada. I thank Martin Green and John Wainright for
helpful discussions. \vskip 0.5 true in

\end{document}